Level of hypnosis Laboratory investigations

**New EEG parameters correlated with sevoflurane concentration in dogs: τ and burst**

Chika Koyama [1,2]*, Taichi Haruna [3], Satoshi Hagihira [4,5], Kazuto Yamashita [1]

[1] Department of Small Animal Clinical Sciences, Graduate School of Veterinary Medicine, Rakuno Gakuen University, Hokkaido, Japan
[2] Department of Anaesthesiology, Japan Animal Referral Medical Center, Tokyo, Japan
[3] Department of Information and Sciences, Tokyo Woman's Christian University, Tokyo, Japan
[4] Department of Anaesthesiology, Kansai Medical University, Osaka, Japan
[5] Department of Anaesthesiology and Intensive Care Medicine, Osaka University Graduate School of Medicine, Osaka, Japan

*Correspondence to Chika Koyama, Department of Small Animal Clinical Sciences, Graduate School of Veterinary Medicine, Rakuno Gakuen University, 582 Midorimachi, Bunkyodai, Ebetsu City, Hokkaido 069-0836, Japan
Address e-mail: chikanesunn@icloud.com

**Abstract**

Background: Electroencephalographic (EEG) indicators for anaesthesia depth are being developed. Here, we propose novel EEG parameters created by using two kinds of waveforms discriminated by voltage thresholds from a new perspective.

Methods: Six young-adult beagles and 6 senior beagles were anaesthetised with end-tidal sevoflurane (SEV) of 2.0%-4.0% in 0.5% intervals, and 2-channel EEG (256 Hz) was recorded. Two events were discriminated: a consecutive part (τ) with a peak-to-peak potential difference (Vpp) within 1-6, 8, 12, or 20 µV, and another part (burst) with Vpp outside the threshold. Number of τ (Nτ), mean τ duration (Mτ), total percentage of τ (SRτ), mean burst duration (Mbst), and amplitude of burst (Abst) were evaluated as anaesthesia depth indicators using Pearson's correlation coefficients (r).

Results: When Vpp was near the suppression wave threshold, Nτ had the highest correlation with SEV in both groups. As SEV was increased until onset of burst suppression, Nτ decreased, Mτ remained unchanged, and Mbst and Abst increased. In the young-adult group, mean |r| exceeded 0.95 for Nτ with Vpp of 4-6 µV, for Mbst with Vpp of 8-12 µV, and for Abst with Vpp of 4-12 µV. In the senior group, r exceeded 0.90 for Nτ with Vpp of 4 µV and for Abst with Vpp of 5-12 µV.

Conclusion: We developed new EEG parameters that were almost perfectly correlated with volatile anaesthetic concentrations using traditional bipolar recording. This study suggests that the effect of anaesthetics on EEG can be investigated from two directions: pacemaker (τ) and generator (burst).

## Introduction

Electroencephalography (EEG) readings are generated mainly by the postsynaptic potential of pyramidal cells in cortical layer 5 (generator) and by functional integration of the ascending reticular activation system and thalamic and cerebral cortical neurons (i.e. pacemaker).[1,2] Since the 1990s, more than 10 EEG-derived monitoring systems have been commercially available.[3,4] However, continuous reports[5-8] of inadequate prevention of intraoperative awareness indicate that accurate assessment of hypnosis level has not yet been achieved.[9,10] Recent research using multichannel EEG has shown that although anaesthesia-induced unconsciousness is related to depressed functional connectivity of brain circuits,[11-13] these connectivity patterns seemed unsuitable for assessing level of consciousness because of their instability.[14]

Here, we introduce a new EEG analysis for developing indicators for depth of anaesthesia. Because a complete suppression wave (i.e., non-electrical activity) is the endpoint of anaesthetic action regardless of species, this study discriminated an ultrashort suppression wave (ms), named "τ" (tau), and a wave between adjacent τ suppression waves, named "burst". The τ waves appear rhythmically on a short time scale. Relationships between sevoflurane concentration and parameters derived from τ and burst were examined using traditional bipolar EEG recording in beagles.

## Methods

### Experimental animals and ethics

The Animal Care and Use Committee of Rakuno Gakuen University approved this study (Approval No. VH14B7) in accordance with "Guidelines for Proper Implementation of Animal Experiments (2006)" by the Science Council of Japan. Six young-adult beagles (2.5 ± 1.5 [mean ± SD] years old; 9.9 ± 0.9 kg) and 6 senior beagles (10.1 ± 1.5 years old; 12.6 ± 1.4 kg) were studied. Each group comprised 3 males and 3 females. Physical examination, complete blood cell count and biochemical analysis confirmed the dogs were healthy. The dogs were cared for according to the principles of the "Guide for the Care and Use of laboratory Animals" prepared by Rakuno Gakuen University.

### Anaesthesia protocols

We have previously reported the detailed protocol for this experiment.[15] The animals were not premedicated. General anaesthesia was induced via mask using 8% sevoflurane in oxygen. After endotracheal intubation, all dogs were placed on mechanical ventilation with sevoflurane and oxygen in left lateral recumbency. Rocuronium bromide was infused (0.5 mg kg$^{-1}$ followed by 1 mg kg$^{-1}$ h$^{-1}$) with Ringer's lactate solution (5 mL kg$^{-1}$ h$^{-1}$) through the cephalic vein. End-tidal $CO_2$ was maintained between 33 and 40 mmHg. Oesophageal temperature was maintained at 37.5-38.0°C using a warm air blanket. The anaesthetic agent monitor was calibrated at the start of each experiment using a calibration kit.

### Data acquisition and preprocessing

EEG data were acquired from Fpz-T4 via needle electrodes placed in a right front-temporal configuration using an A-2000XP BIS monitor® (ver. 3.21, Covidien-Medtronic, Minneapolis, MN).[15] End-tidal sevoflurane concentration (SEV) was maintained in order at 2.0%, 2.5%, 3.0%, 3.5% and 4.0% until reaching a 20-min equilibrium, and the EEG signals were digitised continuously for 5 min. Simultaneously, raw signals were exported into EEG analysis software (Bispectral Analyzer for A2000) and down sampled to 256 Hz.[16] To avoid the checking time of the BIS monitor, 3 EEG data packets of 64 s were extracted for analysis. One-second moving averages were subtracted from the data, which were divided into mutually non-overlapping 2-s periods, and Welch's window function was applied. Subsequently, discrete Fourier transform was performed at 50 ± 1 Hz and 100 ± 1 Hz to remove noise caused by the alternating current power source. Muscle relaxants were used to prevent myoelectric potential noise contamination of the EEG data.

### τ and burst

We first focused on an ultrashort suppression wave. Enlarging the EEG waveform (Figure 1B and C) revealed that the suppression wave was not flat but fluctuating and comprised minute waves with periods of approximately 0.010 s. Such waves ranging from 1 min to a few minutes in duration were scattered throughout the EEG data at any level of anaesthesia, and we refer to these waves as "τ" waves (Figure 1D and E). We defined a waveform between two adjacent τ waves as a "burst". Thus, EEG waveforms were classified into two types of waves, namely τ and burst.

After detecting peaks in EEG data using the first derivative, Each consecutive part in which the potential difference between adjacent peaks (peak-to-peak-voltage) was within a given voltage threshold was defined as τ. Each consecutive part in which the potential difference between adjacent peaks was outside the voltage threshold (between adjacent τ waves) was defined as a burst (Fig. 1A). The potential difference threshold (Vpp) was set at 1-6 µV in 1-µV intervals and additionally at 8, 12 and 20 µV. Despite some individual differences in the measured potential, this measure was roughly classified according to the magnitude of Vpp referenced from the enlarged EEG data as follows. When Vpp was set to approximately 1-3 μV, τ was detected as a microwave. When Vpp was set around 3-8 μV, τ was detected as a fluctuating suppression wave. When Vpp was set to 8-20 μV, a burst was detected as a spike wave.

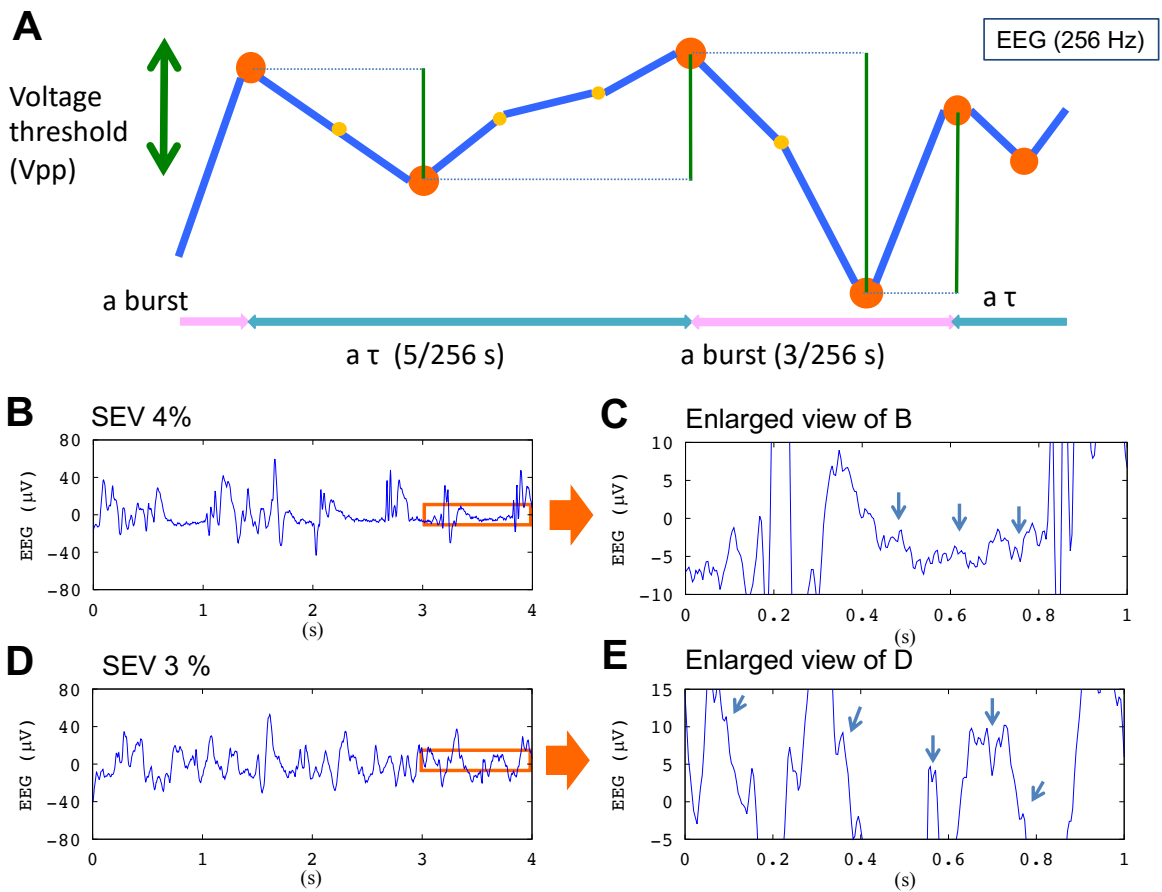

**Figure 1** Concept of "τ" and "burst". (A): Illustration of τ and burst events using a schematic diagram of 256 Hz electroencephalographic (EEG) waveforms. A τ event is defined as a consecutive part on EGG in which the potential difference between adjacent peaks is within a given voltage threshold (Vpp). A burst is defined as a consecutive part in which the potential difference between adjacent peaks exceeds Vpp (i.e. the part between adjacent τ events.) (B): EEG data over 4 s during onset of a burst suppression (BS) under 4% sevoflurane. (C): Enlarged view (1 s, 20 μV) of the red box in (B). A τ event was discriminated based on fluctuating waves of the suppression waveform (arrows). (D): EEG data over 4 s under 3% sevoflurane, which is below the BS level (E): Enlarged view (1 s, 20 μV) of the red box in (D). An ultrashort suppression wave (arrows) is observed occasionally.

### New EEG parameters

Based on τ and burst components in 64 s of EEG data, we developed the following 5 new

EEG parameters:

$\tau$: the consecutive part within the Vpp threshold

$N\tau$ = total number of $\tau$ in 64 s

$M\tau\ (s) = \Sigma\tau\ (s)/N\tau$

$SR\tau = \Sigma\tau\ (s)/64\ (s)$

burst (s): the part between two $\tau$ waves

$Mbst\ (s) = \Sigma burst\ (s)/Nbst$

$Abst\ (\mu V) = \Sigma amp\ (\mu V)/Nbst$

$\Sigma\tau$ (s) denotes the sum of all $\tau$ events. $\Sigma$burst (s) denotes the sum of all burst events. Nbst is the total number of burst events occurring in 64 s and is equal to $N\tau$ or $N\tau \pm 1$. $\Sigma$amp ($\mu$V) is the sum of the difference between the maximum and minimum voltages in a burst, which we call burst amplitude. $M\tau$ (s) is mean $\tau$ duration (s). $SR\tau$ is the total percentage of $\tau$ events in the 64-s interval and can be calculated from $N\tau$ and $M\tau$. Mbst (s) is mean burst duration and can be calculated using $N\tau$ and $M\tau$. Abst ($\mu$V) is mean burst amplitude. Under deep anaesthesia, where burst suppression (BS) appears on EEG, both $M\tau$ and $SR\tau$ within Vpp of 3-8 $\mu$V should increase. Finally, when EEG data comprise a completely flat wave, $N\tau$ equals 1.

Data were analysed by a program written in C. The graphical outputs were generated using gnuplot (Version 5.1, http://www.gnuplot.info/).

Statistical analysis

We investigated statistics summarising $\tau$ and burst in EEG data in beagles anaesthetised with 2.0%-4.0% SEV. The 2.0% concentration was based on the reported value of 1.3 ± 0.3 (%) for the minimum alveolar concentration preventing voluntary response in 50% of dogs (MAC-awake) of SEV.[17] In our previous study using the same EEG data of the senior beagles, we confirmed that 3.5%-4.0% SEV achieved deep anaesthesia with BS evident on EEG.[15]

The strength and sign of the linear relationship between SEV concentration and each parameter ($N\tau$, $M\tau$, $SR\tau$, Mbst and Abst) were analysed at each Vpp value in each dog using Pearson's product-moment correlation coefficient (r). Mean ± SD values of r in each group were calculated at each Vpp. For all analyses, p < 0.05 was considered significant. All statistical evaluations were performed using R (Ver.3.5.1; R Foundation for Statistical Computing, Vienna, Austria).

**Results**

Three results at each SEV were almost constant across all parameters from each animal (Figures 2 and 3). At 3.5% SEV in senior 5 (S5) and at 4.0% in all seniors, where a BS pattern occurred on EEG,[15] both $M\tau$ and $SR\tau$ for Vpp < 8 $\mu$V clearly increased compared to the value at lower

SEV (Figure 3 B0-1, C0-1). Also, in young adult 3 (YA3), both parameters similarly increased compared to the value at lower SEV (Figure 2 B0-1, C0-1). At 3.5% in S5 and at 4.0% in the other seniors and YA3, the other parameters also showed consistently different results from the trends observed at lower SEV. All parameters showed similar trends with respect to SEV changes. Correlations of each parameter with SEV are summarised in Table 1 (young-adult group) and Table 2 (senior group). For all parameters, the r value changed stepwise according to the magnitude of the Vpp value. All |r| values exceeding 0.51 were statistically significant ($p < 0.05$).

Although the Vpp value that maximised r varied among individual animals, Nτ showed strong negative correlation with SEV. In 11 animals other than S3, the highest |r| between SEV and Nτ was 0.93-0.98, including a nearly perfect fit of |r| = 0.98 in 4 animals. Mean Nτ had the strongest negative correlation with SEV for Vpp of 6 µV in the young-adult group and for Vpp of 4 µv in the senior group (Figures 2 and 3). For the same Vpp, Mτ remained within a narrow range (0.02-0.04 s) and was also constant at any SEV <3.5% in each animal. In young-adult animals except YA3 and in S1, S4 and S6, Mbst was strongly correlated with SEV. In YA3, S2, S3 and S5, the correlation between Mbst and SEV weakened as Mbst decreased at 4.0% SEV. In addition, for almost all Vpp values, Abst showed strong correlation with SEV. In 11 animals, except S3, the highest r value between SEV and Abst was 0.91-0.99, including r ≥ 0.97 in 5 animals. Especially in YA1 and YA2, Abst showed remarkably positive correlation (r = 0.99) with SEV for almost all Vpp values.

Table 1 Correlation coefficients between sevoflurane concentration and each parameter, at each voltage threshold (Vpp) in each young-adult beagle (YA1–YA6) and the mean ± SD values in the young-adult group.

|    | Vpp | YA1 | YA2 | YA3 | YA4 | YA5 | YA6 | mean ± SD |
|----|-----|-----|-----|-----|-----|-----|-----|-----------|
| Nτ | 1 | -0.96 | -0.91 | 0.29 | -0.67 | -0.94 | -0.86 | -0.68 ± 0.44 |
|    | 2 | -0.98 | -0.95 | -0.18 | -0.84 | -0.96 | -0.96 | -0.81 ± 0.29 |
|    | 3 | -0.98 | -0.97 | -0.77 | -0.88 | -0.96 | -0.98 | -0.92 ± 0.08 |
|    | 4 | -0.97 | -0.97 | -0.96 | -0.90 | -0.96 | -0.98 | -0.96 ± 0.03 |
|    | 5 | -0.97 | -0.98 | -0.97 | -0.94 | -0.96 | -0.98 | -0.96 ± 0.01 |
|    | 6 | -0.95 | -0.97 | -0.96 | -0.96 | -0.96 | -0.98 | -0.97 ± 0.01 |
|    | 8 | -0.92 | -0.95 | -0.89 | -0.95 | -0.95 | -0.97 | -0.94 ± 0.03 |
|    | 12 | 0.87 | 0.42 | -0.63 | 0.32 | 0.20 | -0.64 | 0.09 ± 0.55 |
|    | 20 | 0.98 | 0.98 | 0.43 | 0.78 | 0.95 | 0.92 | 0.84 ± 0.19 |
| Mτ | 1 | 0.40 | 0.68 | 0.72 | 0.90 | 0.83 | 0.73 | 0.71 ± 0.16 |
|    | 2 | 0.25 | -0.03 | 0.71 | 0.94 | 0.60 | 0.76 | 0.54 ± 0.33 |
|  | 3 | -0.21 | -0.53 | 0.69 | 0.83 | -0.39 | 0.65 | 0.17 ± 0.56 |
|  | 4 | -0.53 | -0.61 | 0.68 | 0.58 | -0.71 | 0.32 | -0.04 ± 0.58 |
|  | 5 | -0.86 | -0.78 | 0.67 | 0.19 | -0.88 | -0.01 | -0.28 ± 0.60 |
|  | 6 | -0.92 | -0.81 | 0.65 | -0.16 | -0.91 | -0.41 | -0.43 ± 0.56 |
|  | 8 | -0.98 | -0.93 | 0.60 | -0.52 | -0.90 | -0.77 | -0.58 ± 0.55 |
|  | 12 | -0.97 | -0.97 | 0.41 | -0.70 | -0.93 | -0.91 | -0.68 ± 0.50 |
|  | 20 | -0.94 | -0.94 | -0.45 | -0.77 | -0.95 | -0.91 | -0.83 ± 0.18 |
| SRτ | 1 | -0.93 | -0.86 | 0.54 | -0.25 | -0.92 | -0.72 | -0.52 ± 0.53 |
|  | 2 | -0.97 | -0.92 | 0.55 | -0.63 | -0.95 | -0.86 | -0.63 ± 0.54 |
|  | 3 | -0.98 | -0.94 | 0.53 | -0.72 | -0.95 | -0.91 | -0.66 ± 0.54 |
|  | 4 | -0.98 | -0.96 | 0.51 | -0.76 | -0.96 | -0.95 | -0.68 ± 0.54 |
|  | 5 | -0.98 | -0.97 | 0.46 | -0.80 | -0.96 | -0.95 | -0.70 ± 0.52 |
|  | 6 | -0.98 | -0.97 | 0.42 | -0.83 | -0.96 | -0.94 | -0.71 ± 0.51 |
|  | 8 | -0.98 | -0.99 | 0.32 | -0.83 | -0.96 | -0.95 | -0.73 ± 0.47 |
|  | 12 | -0.98 | -0.99 | 0.08 | -0.86 | -0.98 | -0.96 | -0.78 ± 0.39 |
|  | 20 | -0.97 | -0.99 | -0.50 | -0.89 | -0.98 | -0.94 | -0.88 ± 0.17 |
| Mbst | 1 | 0.95 | 0.88 | -0.15 | 0.65 | 0.93 | 0.85 | 0.68 ± 0.39 |
|  | 2 | 0.98 | 0.93 | 0.03 | 0.84 | 0.95 | 0.94 | 0.78 ± 0.34 |
|  | 3 | 0.98 | 0.96 | 0.27 | 0.89 | 0.95 | 0.96 | 0.83 ± 0.25 |
|  | 4 | 0.98 | 0.97 | 0.55 | 0.90 | 0.95 | 0.96 | 0.89 ± 0.15 |
|  | 5 | 0.98 | 0.97 | 0.76 | 0.93 | 0.96 | 0.97 | 0.93 ± 0.08 |
|  | 6 | 0.98 | 0.98 | 0.82 | 0.94 | 0.96 | 0.97 | 0.94 ± 0.05 |
|  | 8 | 0.97 | 0.98 | 0.87 | 0.95 | 0.97 | 0.97 | 0.95 ± 0.04 |
|  | 12 | 0.94 | 0.98 | 0.85 | 0.97 | 0.97 | 0.96 | 0.95 ± 0.04 |
|  | 20 | 0.69 | 0.96 | 0.42 | 0.95 | 0.90 | 0.89 | 0.80 ± 0.19 |
| Abst | 1 | 0.99 | 0.97 | 0.15 | 0.75 | 0.96 | 0.91 | 0.79 ± 0.29 |
|  | 2 | 0.99 | 0.98 | 0.44 | 0.84 | 0.96 | 0.94 | 0.86 ± 0.19 |
|  | 3 | 0.99 | 0.98 | 0.75 | 0.88 | 0.97 | 0.95 | 0.92 ± 0.08 |
|  | 4 | 0.99 | 0.99 | 0.93 | 0.89 | 0.97 | 0.95 | 0.96 ± 0.03 |
|  | 5 | 0.99 | 0.99 | 0.97 | 0.91 | 0.97 | 0.94 | 0.96 ± 0.03 |
|  | 6 | 0.99 | 0.99 | 0.97 | 0.92 | 0.97 | 0.95 | 0.96 ± 0.03 |
|  | 8 | 0.99 | 0.99 | 0.95 | 0.93 | 0.98 | 0.94 | 0.96 ± 0.02 |
|  | 12 | 0.99 | 0.99 | 0.93 | 0.95 | 0.98 | 0.91 | 0.96 ± 0.03 |
|  | 20 | 0.97 | 0.98 | 0.92 | 0.96 | 0.97 | 0.85 | 0.94 ± 0.05 |

$|r| > 0.51$ was statistically significant ($p < 0.05$).

Table 2 Correlation coefficients between sevoflurane concentration and each parameter at each voltage threshold (Vpp) in each senior beagle (S1–S6) and the mean ± SD values in the senior group.

|     | Vpp | S1    | S2    | S3    | S4    | S5    | S6    | mean ± SD     |
| --- | --- | ----- | ----- | ----- | ----- | ----- | ----- | ------------- |
| Nτ  | 1   | -0.04 | 0.31  | -0.58 | -0.87 | 0.70  | -0.94 | -0.28 ± 0.66  |
|     | 2   | -0.95 | -0.12 | -0.68 | -0.93 | -0.94 | -0.98 | -0.76 ± 0.31  |
|     | 3   | -0.98 | -0.74 | -0.76 | -0.90 | -0.95 | -0.97 | -0.88 ± 0.10  |
|     | 4   | -0.98 | -0.93 | -0.73 | -0.86 | -0.92 | -0.97 | -0.90 ± 0.08  |
|     | 5   | -0.96 | -0.95 | -0.59 | -0.86 | -0.90 | -0.95 | -0.87 ± 0.13  |
|     | 6   | -0.95 | -0.95 | -0.38 | -0.78 | -0.89 | -0.91 | -0.81 ± 0.13  |
|     | 8   | -0.92 | -0.92 | 0.23  | -0.65 | -0.86 | -0.62 | -0.62 ± 0.40  |
|     | 12  | -0.77 | -0.72 | 0.71  | 0.36  | -0.78 | 0.65  | -0.09 ± 0.67  |
|     | 20  | 0.88  | 0.46  | 0.88  | 0.91  | -0.43 | 0.94  | 0.61 ± 0.49   |
| Mτ  | 1   | 0.90  | 0.87  | 0.57  | 0.77  | 0.87  | 0.88  | 0.81 ± 0.11   |
|     | 2   | 0.85  | 0.88  | 0.62  | 0.72  | 0.82  | 0.81  | 0.78 ± 0.09   |
|     | 3   | 0.84  | 0.86  | 0.59  | 0.71  | 0.79  | 0.75  | 0.76 ± 0.09   |
|     | 4   | 0.82  | 0.84  | 0.50  | 0.69  | 0.76  | 0.66  | 0.71 ± 0.11   |
|     | 5   | 0.78  | 0.81  | 0.35  | 0.69  | 0.72  | 0.51  | 0.64 ± 0.16   |
|     | 6   | 0.77  | 0.79  | 0.16  | 0.55  | 0.71  | 0.35  | 0.55 ± 0.23   |
|     | 8   | 0.71  | 0.74  | -0.31 | 0.36  | 0.69  | -0.25 | 0.32 ± 0.45   |
|     | 12  | 0.31  | 0.55  | -0.68 | -0.65 | 0.66  | -0.79 | -0.10 ± 0.62  |
|     | 20  | -0.90 | -0.49 | -0.85 | -0.86 | 0.54  | -0.94 | -0.58 ± 0.53  |
| SRτ | 1   | 0.76  | 0.67  | 0.10  | 0.66  | 0.79  | 0.25  | 0.54 ± 0.26   |
|     | 2   | 0.72  | 0.70  | 0.25  | 0.59  | 0.79  | -0.12 | 0.49 ± 0.32   |
|     | 3   | 0.69  | 0.69  | 0.19  | 0.55  | 0.78  | -0.38 | 0.42 ± 0.40   |
|     | 4   | 0.61  | 0.68  | 0.08  | 0.48  | 0.77  | -0.52 | 0.35 ± 0.45   |
|     | 5   | 0.49  | 0.64  | -0.04 | 0.41  | 0.76  | -0.67 | 0.27 ± 0.49   |
|     | 6   | 0.42  | 0.62  | -0.16 | 0.09  | 0.74  | -0.72 | 0.16 ± 0.50   |
|     | 8   | 0.15  | 0.53  | -0.38 | -0.32 | 0.70  | -0.83 | -0.02 ± 0.54  |
|     | 12  | -0.49 | 0.28  | -0.62 | -0.83 | 0.62  | -0.89 | -0.32 ± 0.57  |
|     | 20  | -0.90 | -0.48 | -0.80 | -0.83 | 0.26  | -0.89 | -0.61 ± 0.42  |
| Mbst | 1  | -0.14 | -0.29 | 0.49  | 0.32  | -0.73 | 0.92  | 0.26 ± 0.43   |
|  | 2 | 0.70 | -0.09 | 0.50 | 0.96 | -0.44 | 0.97 | 0.44 ± 0.53 |
|  | 3 | 0.89 | 0.27 | 0.60 | 0.94 | -0.20 | 0.98 | 0.58 ± 0.42 |
|  | 4 | 0.92 | 0.52 | 0.64 | 0.92 | -0.05 | 0.98 | 0.65 ± 0.36 |
|  | 5 | 0.93 | 0.61 | 0.62 | 0.93 | 0.16 | 0.98 | 0.70 ± 0.29 |
|  | 6 | 0.93 | 0.65 | 0.52 | 0.95 | 0.27 | 0.96 | 0.71 ± 0.26 |
|  | 8 | 0.94 | 0.69 | 0.41 | 0.95 | 0.30 | 0.94 | 0.70 ± 0.27 |
|  | 12 | 0.94 | 0.62 | -0.29 | 0.92 | -0.03 | 0.86 | 0.50 ± 0.49 |
|  | 20 | 0.87 | 0.16 | -0.81 | 0.43 | -0.42 | 0.04 | -0.03 ± 0.57 |
| Abst | 1 | 0.53 | 0.01 | 0.75 | 0.57 | -0.72 | 0.93 | 0.35 ± 0.55 |
|  | 2 | 0.93 | 0.44 | 0.87 | 0.91 | -0.39 | 0.94 | 0.62 ± 0.48 |
|  | 3 | 0.96 | 0.84 | 0.89 | 0.88 | 0.02 | 0.94 | 0.76 ± 0.33 |
|  | 4 | 0.96 | 0.95 | 0.89 | 0.86 | 0.55 | 0.94 | 0.86 ± 0.14 |
|  | 5 | 0.96 | 0.97 | 0.88 | 0.87 | 0.93 | 0.93 | 0.92 ± 0.04 |
|  | 6 | 0.96 | 0.97 | 0.87 | 0.85 | 0.93 | 0.92 | 0.92 ± 0.04 |
|  | 8 | 0.95 | 0.98 | 0.88 | 0.85 | 0.91 | 0.92 | 0.92 ± 0.04 |
|  | 12 | 0.96 | 0.96 | 0.88 | 0.88 | 0.89 | 0.92 | 0.91 ± 0.03 |
|  | 20 | 0.97 | 0.93 | 0.89 | 0.81 | 0.85 | 0.85 | 0.88 ± 0.05 |

$|r| > 0.51$ was statistically significant ($p < 0.05$).

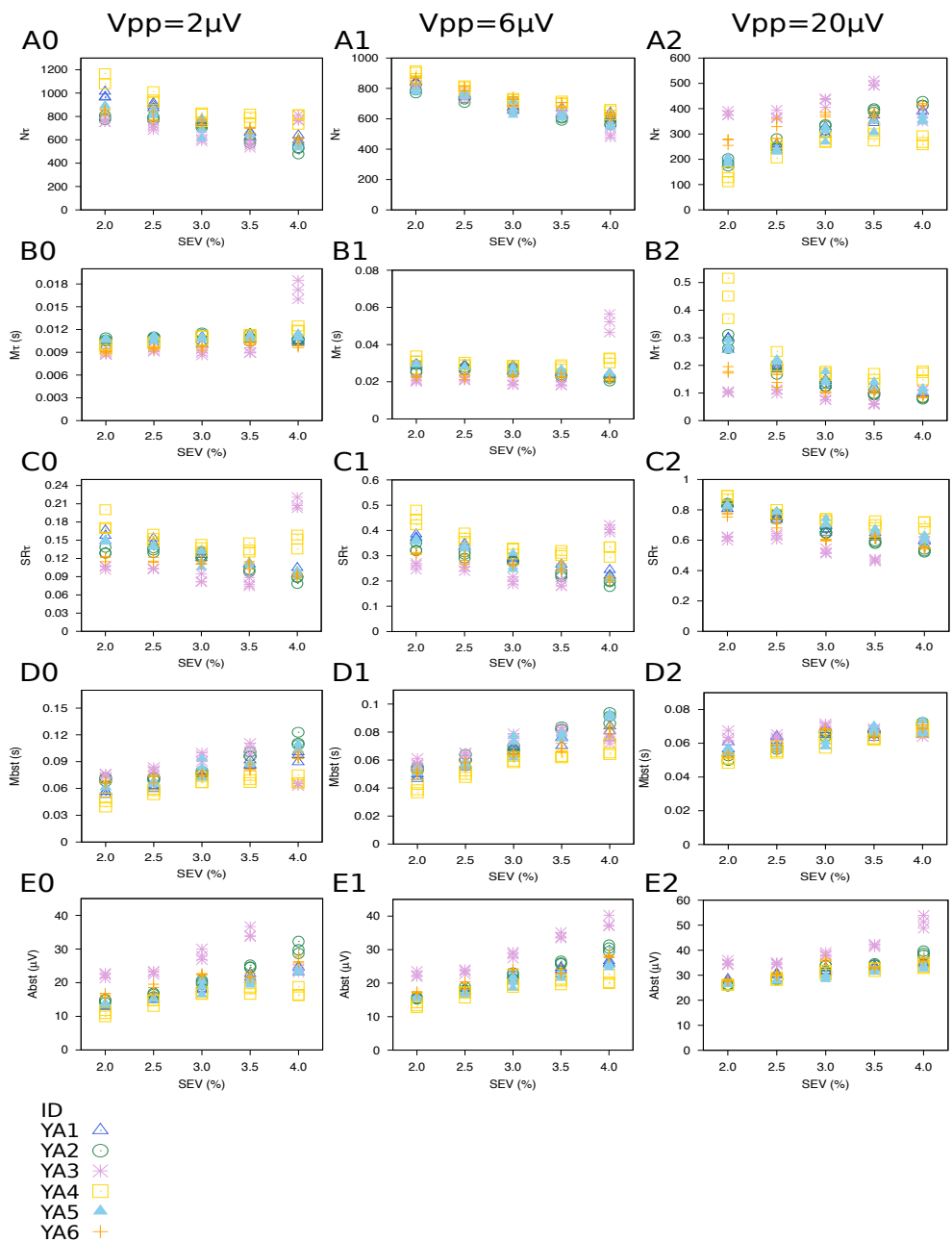

**Figure 2** Examples of the relationship between end-tidal sevoflurane concentration (SEV) and the newly proposed electroencephalographic (EEG) parameters Nτ, Mτ, SRτ, Mbst and Abst calculated with thresholds of potential difference (Vpp) of 2, 6 and 20 μV in 6 young-adult beagles. In each animal, three EEG datasets of 64 s were analysed at each SEV. For Vpp of 2 μV compared with 1–20 μV (excluding 2 μV), the number of τ events (Nτ) was the highest, such that τ would detect a microwave. For Vpp of 6 μV compared with 1–20 μV (excluding 6 μV), mean Nτ had the strongest negative correlation with SEV, where τ would detect an ultrashort suppression wave. For Vpp of 20 μV, a burst would detect a spike. The number of bursts is equal to $N\tau$ or $N\tau \pm 1$.

A: $N\tau$ in 64 s vs SEV. (A0): $N\tau$ decreased with increasing SEV and increased when burst suppression (BS) appeared on EEG. (A1): $N\tau$ decreased with increasing SEV, with low variation among individual dogs. (A2): $N\tau$ (= number of bursts) increased with increasing SEV. B: Mean $\tau$ duration ($M\tau$) vs SEV. (B1): $M\tau$ was almost constant between 0.02 s and 0.04 s in all animals except YA3 at 4.0% SEV, where the EEG data included a BS pattern. (B2): Mt was the highest at 2.0% SEV, where EEG data contained some low-amplitude fast waves. (C): Total percentage of $\tau$ in 64 s ($SR\tau$) vs SEV. $SR\tau = 64 - N\tau \times M\tau$. (D): Mean burst duration (Mbst) vs SEV. Mbst = $64/ N\tau - M\tau$. (E): Mean potential difference between the maximum and minimum values of peaks (Abst) vs SEV. (E1) Abst had a strong correlation ($r \geq 0.97$, n = 4; r = 0.95, n = 1; r = 0.92, n =1) with SEV in all animals.

12208c498461e7a4829d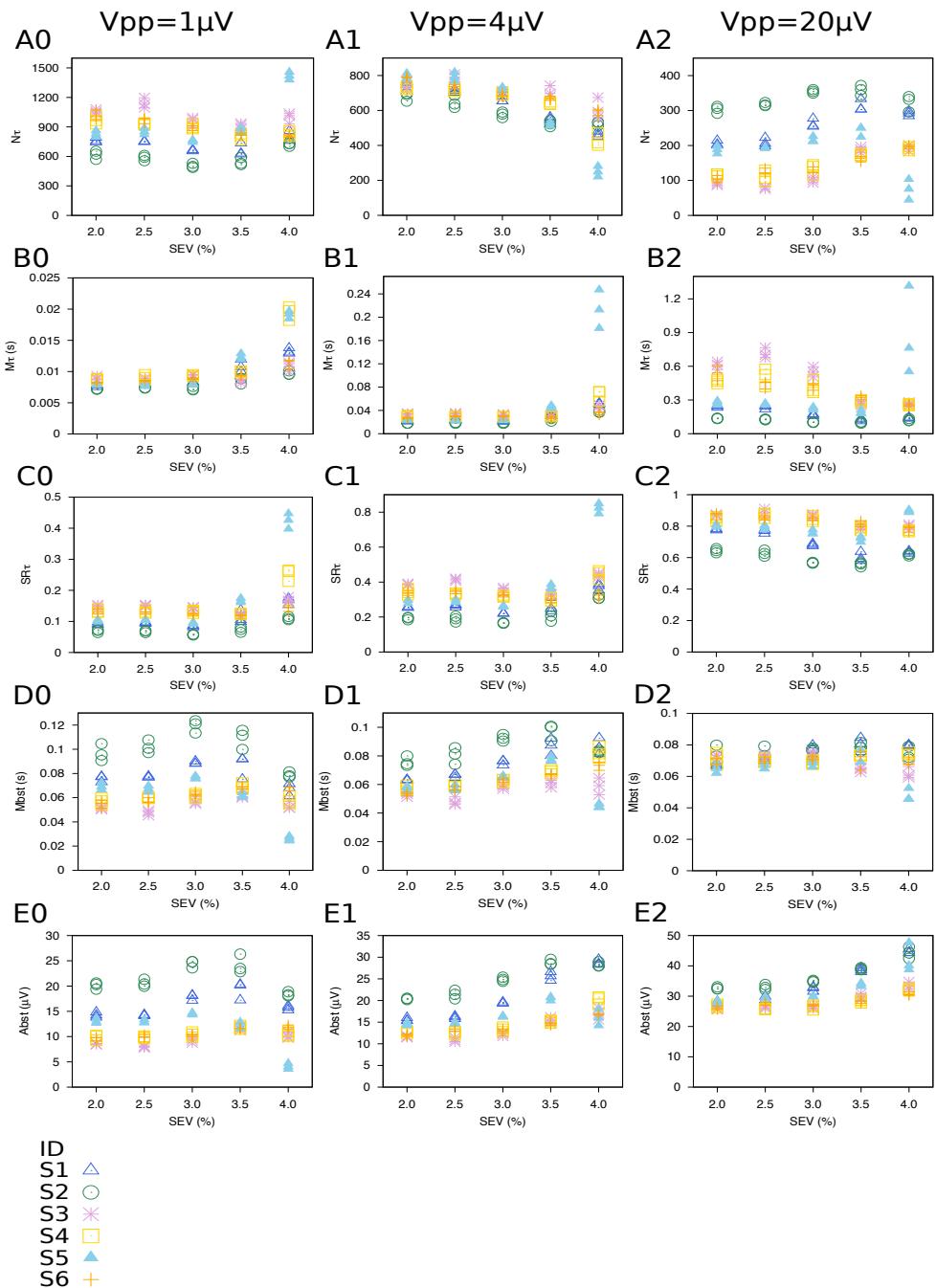

**Figure 3** Examples of the relationship between end-tidal sevoflurane concentration (SEV) and the newly proposed electroencephalographic (EEG) parameters Nτ, Mτ, SRτ, Mbst and Abst calculated with potential difference (Vpp) thresholds of 1, 4 and 20 μV in 6 senior beagles. In each animal, three EEG datasets of 64 s were analysed at each SEV. For Vpp of 1 μV compared with 1–20 μV (excluding 1 μV), the number of τ events (Nτ) was the highest, such that τ would detect a microwave. For Vpp of 4 μV compared with 1–20 μV (excluding 4 μV), Nτ had the strongest negative correlation with SEV, such that τ would detect an ultrashort

suppression wave. For Vpp of 20 µV, a burst would detect a spike. The number of bursts is equal to $N\tau$ or $N\tau \pm 1$.

A: $N\tau$ in 64 s vs SEV. (A0): $N\tau$ decreased with SEV and increased when burst suppression (BS) appeared on EEG. (A1): $N\tau$ decreased with increasing SEV, with low variation among individual dogs. At 3.5–4.0% SEV, $N\tau$ decreased greatly depending on the degree of the suppression ratio. (B): Mean $\tau$ duration ($M\tau$) vs SEV. (B1): $M\tau$ was almost constant between 0.02 s and 0.04 s from 2.0% to 3.5% SEV in all animals except in S5 at 3.5% SEV, where the EEG data included a BS pattern. (C): Total percentage of $\tau$ in 64 s ($SR\tau$) vs SEV: $SR\tau = 64 - N\tau \times M\tau$. $SR\tau$ and $M\tau$ increased when BS appeared on EEG. (D): mean burst (Mbst) vs SEV: $Mbst = 64/N\tau - M\tau$. Mbst showed a tendency similar to Abst. (E): Mean potential difference between the maximum value and the minimum value of peaks (Abst) vs SEV. (E1) Abst had a strong correlation ($r > 0.95$, $n = 2$; $r = 0.94$, $n = 1$; $r > 0.85$, $n = 2$; $r = 0.55$, $n = 1$) with SEV in all animals except S5, where the EEG data were almost completely occupied by suppression.

**Discussion**

Here we have presented a new EEG analysis method based on two kinds of waves originally developed in this study. One wave consisting of the potential difference between adjacent peaks within a threshold of voltage (Vpp) was defined as a τ event, and the other wave exceeding the threshold was defined as a burst. Increasing the value of Vpp changes target detection to microwave, suppression wave and spike wave events. The parameters created based on τ and burst events showed specific trends with respect to depth of anaesthesia in all animals. The number of τ events (Nτ) showed almost perfect negative correlation with SEV in almost all animals when Vpp was 2−6 μV. For those Vpp values, mean τ duration of (Mτ) was 0.02-0.04 s, with small individual differences, and was constant at any SEV in the absence of BS on EEG. When BS was clearly present on EEG, Mτ and the total percentage of τ (SRτ) increased as expected. Mean burst duration (Mbst) increased with increasing SEV in the absence of BS. The amplitude of burst (Abst) showed almost perfect positive correlation with SEV across a wide range of Vpp values in all animals. Additionally, the correlation coefficient between those parameters and SEV changed stepwise according to the magnitude of Vpp. To our knowledge, this is the first report to introduce EEG parameters having a strong linear correlation close to ±1 with volatile anaesthetic concentration.

In our previous study, enlarged EEG waveforms were observed to measure the suppression ratio, and it seemed that an ultrashort suppression appeared occasionally but rhythmically at any SEV. Moreover, a complete suppression wave (Nτ = 1) is the endpoint of anaesthetic action regardless of species. Hence, in this study, an ultrashort suppression wave (Figure 1E) was discriminated as a τ event to investigate depth of anaesthesia. The frequency of τ events showed very strong negative correlation with anaesthetic concentration. In addition, our previous study also showed that there was value in focusing on the ultrashort suppression wave. Suppression wave was defined as <2.25 μV using the same EEG data of the senior dogs.[15] This value corresponds to Vpp of 4.5 μV, which is close to 4 μV, which occurred when |r| was highest in the senior group. Furthermore, the behaviour of Nτ is consistent with previous findings comparing young-adult and senior beagles. In general, the amount of anaesthesia required decreases with age,[18][19] and the measured potential is higher at a younger age, with less resistance because of more moisture and less fibre. Indeed, for example, mean Nτ for Vpp of 6 μV in the young-adult group, which was most strongly correlated with SEV, was greater than the Vpp of 4 μV in the senior group at any SEV (Nτ = 829 vs 748, Nτ = 766 vs 731, Nτ = 692 vs 675, Nτ = 653 vs 601, Nτ = 585 vs 478 at 2.0%, 2.5%, 3.0%, 3.5%, and 4.0% SEV, respectively). Three key conclusions can be made based on these findings. (1) The strongest correlation between the number of τ events and SEV was when Vpp was close to the suppression criterion. (2) In almost all animals, the number of τ events correlated fairly strongly with SEV. (3) The Nτ results were consistent with age differences in depth of hypnosis and measurement of potential. These results suggest that τ is a suitable anaesthetic indicator. Additionally, τ, similar to some fluctuations in suppression waves (Figure 1E), seems to show a momentary electrical pause synchronised in some areas, resembling a rhythm between electrical activity. In that case, a decrease in Nτ means that the tempo of the electrical pauses slows as anaesthesia deepens. Furthermore, BS, an indicator of excessive anaesthesia, tended to appear when Nτ was <600 in this study, which we presume depended on the amount of nerve tissue monitored by the electrodes.

The thalamus generates rhythmic patterns in EEG data.[20][21] Anaesthetic concentrations correlate with rhythmic EEG represented by spectral edge frequency 95 (SEF95) as analysed by Fourier transform.[22][23] EEG data represent surface potentials primarily reflecting synaptic activity in dendrites of cortical pyramidal neurons, which have mutual fibre communications with thalamic neurons. This circuit is monitored by the reticular formation in the brainstem using acetylcholine as a transmitter. Rhythmic firing of thalamic neurons produces excitatory postsynaptic potentials (EPSP) synchronously with pyramidal neurons. That frequency is determined by EPSP duration and inhibitory postsynaptic potential (IPSP) duration in thalamic neurons. In other words, it is dependent on the membrane potential level of the thalamus.[21][24][25] Anaesthetics, including sevoflurane, act on γ-aminobutyric acid A receptors (GABA$_A$) to open chloride channels and inhibit EPSP transmission and promote IPSP transmission.[26][27] Furthermore, volatile anaesthetics have an inhibitory effect on the hyperpolarisation activated current (H current) that slowly causes depolarisation of post-hyperpolarisation and contributes to the generation of synchronised oscillatory neural rhythms.[28][29] Indeed, use of 1%-4% sevoflurane dose-dependently inhibits H current.[30] The inhibition of H current may lead to tempo delay.

In this study, Abst showed strong linear correlation close to 1 with SEV for almost all Vpp values. In contrast, although Mbst calculated from Nτ and Mτ was nearly proportional to Abst in the absence of BS, the proportional relationship no longer held at BS levels (supplementary). These results might reflect the effect of anaesthetics on the postsynaptic potential activity of pyramidal cells in the neocortex, the main component of EEG readings. High-amplitude slow waves on EEG due to the administration of anaesthetics are well known. Through the hyperpolarisation of thalamic cortical neurons, delta waves are generated on EEG.[25] Moreover, during BS, the hyperpolarisation of cortical neurons blocks input from the thalamus, which then stops working. Endogenous pacemakers can activate 30%-40% of thalamic neurons, inducing activity of thalamic cortical neurons. They also activate cortical neurons from the hyperpolarised membrane potential as burst waves in the BS pattern.[31][32] In this study, we defined Abst as the potential difference indicating the magnitude of current energy. When deciding which EEG parameters to set, both the root mean square (RMS) value of amplitude and Abst were examined. Although the relationship between SEV and RMS was similar to that of Abst, the r value was inferior.

In addition to the near-perfect correlation between SEV and Abst across a wide range of Vpp values, the r value between SEV and Nτ changed stepwise according to Vpp magnitude. As the threshold was increased, the duration of detection increased and the number of detections decreased, in a manner that was dependent on the distribution of the amplitude of the EEG waveform. Accordingly, although it is difficult to accurately detect small potential differences of suppression waves owing to individual differences in resistance based on tissue and skeleton, a value close to the actual number of ultrashort suppressions can be detected with an approximate Vpp threshold. Even for a large difference in amplitude between individuals, Vpp can be set each time based on Mτ, which has an almost constant value and lacks individual differences. Also, this method uses a popular 2-channel bipolar recording method. These results show that such new parameters could be widely applicable.

We assumed that EEG waveforms show BS patterns when both Mτ and SRτ increase. This can be confirmed from EEG waveform data by visual inspection and our previous results using the same EEG data of the senior group in this study. As a reference, the mean suppression ratio (SR) for 64 s in three sets of EEG data in S1, S2, S3, S4, S5 and S6, were 23.3%, 20.0%, 14.2%, 25.7%, 85.9%, and 7.0% at 4.0% SEV; 8.4%, 3.4%, 0%, 0.4%, 19.0% and 0.4% at 3.5% SEV; and uniformly 0% for all results at 2.0%-3.0% SEV, respectively. With SR <10% in S1 and S2, both Mτ and SRτ showed no noticeable change at 3.5%. A slight decrease in Mτ as SEV increases may compensate for this. Above all, in all animals, Mτ remained within a narrow range (0.02-0.04 s) for all SEV concentrations in the absence of BS. Through further research, we would like to elucidate the significance of this interesting phenomenon, taking into account the anaesthetic action or consciousness associated with a discrete event.

In conclusion, we have developed a new EEG analysis method focusing on an ultrashort suppression wave with duration of milliseconds, which we named "τ", and an ultrashort wave between adjacent τ events, which we named "burst". Both τ and burst events showed strong correlation with SEV. Furthermore, these findings might explain the mechanism of EEG generation from two different perspectives, considering τ events as "pace-makers" and burst events as "generators".

Acknowledgement: We gratefully acknowledge Jun Tamura and the students who supported this experiment at Rakuno Gakuen University, and Masanori Murayama and his colleagues at RIKEN Center for Brain Science for providing an opportunity for two-photon imaging of the somatosensory cortex that supported the interpretation of the new EEG parameters.

Conflicts of interest: None

Author's contribution

C.K. contributed to study design, proposal of the new parameters, data collection and analysis and drafting of the manuscript; T.H. contributed to creation of the analysis program, data analysis and revision of the manuscript; S.H. contributed to study design and revision of the manuscript; K.Y. contributed experimental design, data collection and revision of the manuscript.

703-11